# Ignition of Deflagration and Detonation Ahead of the Flame due to Radiative Preheating of Suspended Micro Particles

M.F. Ivanov [1], A. D. Kiverin [1], M. A. Liberman [2,*]

[1] Joint Institute for High Temperatures, Russian Academy of Science, Moscow 125412, Russia

[2] Nordita, KTH Royal Institute of Technology and Stockholm University Roslagstullsbacken 23, SE 10691 Stockholm, Sweden

## Abstract

We study a flame propagating in the gaseous combustible mixture with suspended inert solid micro particles. The gaseous mixture is assumed to be transparent for the radiation emitted by the combustion products, while particles absorb and re-emit the radiation. Thermal radiation heats the particles, which in turn transfer the heat to the surrounding unburned gaseous mixture by means of heat conduction, so that the gas phase temperature lags that of the particles. We consider different scenarios depending on the spatial distribution of the particles, their size and the number density. In the case of uniform spatial distribution of the particles the radiation causes a modest increase of the temperature ahead of the flame and corresponding modest increase of the combustion velocity. The effects of radiation preheating as stronger as smaller the normal flame velocity. On the contrary, in the case of non-uniform distribution of the particles, such that the particles number density is relatively small in the region just ahead of the flame front and increases in the distant region ahead of the flame, the preheating caused by the thermal radiation absorption may trigger additional independent source of ignition. This scenario requires the formation of a temperature gradient with the maximum temperature sufficient for ignition in the region of denser particles cloud ahead of the advancing flame. Depending on the steepness of the temperature gradient formed in the unburned mixture, either deflagration or detonation can be initiated via the Zeldovich's gradient mechanism. The ignition and the resulting combustion regimes depend on the number density profile and, correspondingly, on the temperature profile which is formed in effect of radiation absorption and gas-dynamic expansion. In the case of coal dust flames propagating through a layered dust cloud the effect of radiation heat transfer can result in the propagation of combustion wave with velocity up to 1000m/s and can be a plausible explanation of the origin of dust explosion in coal mines.

___________________________________________
*Corresponding author. E-mail: michael.liberman@nordita.org

# 1. Introduction

It is known, that uncontrolled development of detonation poses significant threats to chemical storage and processing facilities, mining operations, etc. [1, 2], while controlled detonation initiation can be a potential application for propulsion and power devices [3]. In astrophysics type Ia Supernovae (SN Ia) were used as a standard distance indicators and led to the one of the most impressive discovery of the accelerating expansion of the universe [4-6]. However, the nature of thermonuclear explosion of SN Ia and specifically the mechanism of transition to detonation during the thermonuclear explosion of SN Ia still remains the least understood aspect of the thermonuclear Supernovae explosion phenomenon [7-11].

While studying combustion in gaseous mixture, the radiation of hot combustion products is not taken into account, as the radiation absorption length in a gaseous mixture is very large, so that the gaseous mixture is almost fully transparent for the radiation and therefore the radiation heat transfer does not influence the flame dynamics. The situation changes drastically in the presence of particulates of size $r_p$ suspended in the gaseous mixture, which is typical for e.g., coal mine, chemical industry, forest fire, etc. In such a case the absorption length of radiation can be estimated as $L = 1/N_p \sigma_p$, where $N_p$ is the particles number density, and $\sigma_p = \pi r_p^2$ is the particles cross section for the light absorption. The radiant energy flux is absorbed by the particles and then lost by conduction from the particles surface to the surrounding unreacted gaseous phase so that the gas phase temperature lags that of the particles. It should also be noted that even the presence of a relatively small concentration of particles increases the luminosity considerably, so that the radiant flux emitted by the hot combustion products may be well approximated by the black-body radiation. Particles preheated by the absorbed radiation increase temperature of the surrounding unburned gaseous mixture affecting the flame dynamics. Studies of the premixed flames and detonations arising and propagating in the presence of suspended particulates may play an important role for the understanding of unconfined vapor cloud explosions, accidental explosions in the coal mines and in the chemical industry, dust explosion hazards, and for better performance of rocket engines using the solid fuels. In the present paper we show that the radiative preheating may not only affect the flame structure, increase the velocity of the propagating flame, but also can trigger a new source of ignition of either deflagration or detonation in the unburned mixture ahead of the flame front.

A combustible mixture can be ignited by electrical sparks, or by thermal heating. The ignition capability of an electrical spark varies with fuel concentration, humidity, oxygen content of the atmosphere, temperature, and turbulence, requiring about 0.01-0.03mJ depending on the mixture reactivity. In contrast, radiation-induced ignition typically requires much larger amounts of energy to be released in the mixture. Direct thermal ignition of gaseous combustible mixture by absorption of radiation causing a rapid increase in temperature at least up to 1000K is possible by focusing a high power laser radiation and has been demonstrated both theoretically and experimentally [12, 13]. However, ignition at low power levels is unlikely because of a very large length of absorption of the combustible gases at normal conditions.

The flame propagating in the uniformly dispersed, quiescent, gravity-free, particle clouds and radiation affected combustion in the presence of inert particles has been studied by different groups of authors assuming uniform dispersion of particulates with and without account of radiative heat transfer [14-22]. The dynamics of particle-laden flames affected by the radiative preheating has been studied in [14-16] using asymptotic methods and a one-step Arrhenius chemical model with high activation energy. Coal combustion research [22, 23] is typically focused on two aspects of practical interest: production of volatiles due to thermal decomposition of coal dust and char combustion. For the coal-dust suspension in air filling the coal-fired burners and rocket engines using the solid fuels as well as for coal-fire mining safety problems both the ignition and combustion evolution are of paramount importance [22, 23]. The combustible volatiles may essentially contribute to the heat-up of the coal particles, enhance the combustion energy release due to energy feedback mechanism resulting in an explosion. Effect of the radiation heat transfer on a spray combustion, which can be of interest for practical cases such as diesel engines, gas turbine combustors etc. was studied in [24].

In practice, the plausibility of ignition is determined by the ignition conditions implying certain energy input with a certain rate. As the mechanism formulated above involves suspended particles as a carrier of radiation energy and since particles have to be heated up to relatively high temperature, their spatial distribution should be such, that the radiation will be absorbed mostly far ahead of the flame front in order to promote the ignition conditions before the flame arrival. Ignition may occur via the Zeldovich temperature gradient mechanism [25] with initiating of one or another combustion regime depending on the gradient steepness. The thermal energy accumulated by the particles and transferred to the gaseous mixture depends on the radiant energy absorbed by the particles, the energy transferred from the particle surface to surrounding gaseous mixture and the radiant loss from the particles.

In the present paper we consider the effects of thermal radiative preheating using as an representative example a flame propagating in the suspension comprising two phases: hydrogen oxygen gaseous mixture and inert solid micro particles. The hydrogen oxygen gaseous mixture is assumed to be transparent for radiation, while the solid particles absorb and reemit the radiation. We consider different scenarios depending on spatial distribution of the suspended particles ahead of the propagating flame. For a uniform spatial distribution of suspended particles in the gaseous mixture the radiative preheating of the particles ahead of the flame results in the increase of the gaseous mixture temperature and correspondingly in the increase of the combustion wave velocity. All the same the effect of radiation is noticeably stronger for the particle-dust flames with a small normal velocity, for example, for methane-air coal dust flames. On the contrary, a non-uniform spatial distribution of dispersed particles, such that the particles concentration is smaller in immediate proximity ahead the flame and increases ahead of the flame forming a denser cloud of particles suspended in the gaseous mixture, may result in ignition of either deflagration or detonation via the Zeldovich gradient mechanism. In the case of coal dust flames propagating through the layered particle-gas cloud the effect of radiation heat transfer can result in the spread of combustion wave with velocity up to 1000m/s and presumably explain the origin of dust explosion in coal mines.

The paper is organized as follows. Section 2 is the formulation of the problem and short description of the numerical method. In Section 3 we perform 1D direct numerical simulations and consider the planar hydrogen-oxygen flame propagating through the mixture with uniform distribution of small suspended particles. Section 4 presents analysis of the ignition of different combustion regimes initiated by the radiative preheating in the case of non-uniform distribution of dispersed particles. We conclude in Section 5. Details of the numerical scheme used in simulations together with thorough convergence and resolution tests are presented in Appendixes A and B.

**2. Governing equations**

The governing equations for a planar flame in the gaseous phase are the one-dimensional, time-dependent, multispecies reactive Navier-Stokes equations including the effects of compressibility, molecular diffusion, thermal conduction, viscosity, chemical kinetics and the chemical energy release, momentum and heat transfer between the particles and the gas:

$$\frac{\partial \rho}{\partial t} + \frac{\partial (\rho u)}{\partial x} = 0, \tag{1}$$

$$\frac{\partial Y_i}{\partial t} + u\frac{\partial Y_i}{\partial x} = \frac{1}{\rho}\frac{\partial}{\partial x}\left(\rho D_i \frac{\partial Y_i}{\partial x}\right) + \left(\frac{\partial Y_i}{\partial t}\right)_{ch}, \tag{2}$$

$$\rho\left(\frac{\partial u}{\partial t} + u\frac{\partial u}{\partial x}\right) = -\frac{\partial P}{\partial x} + \frac{\partial \sigma_{xx}}{\partial x} - \rho_p \frac{(u-u_p)}{\tau_{St}}, \tag{3}$$

$$\rho\left(\frac{\partial E}{\partial t} + u\frac{\partial E}{\partial x}\right) = -\frac{\partial Pu}{\partial x} + \frac{\partial}{\partial x}(\sigma_{xx} u) + \frac{\partial}{\partial x}\left(\kappa(T)\frac{\partial T}{\partial x}\right) +$$

$$+ \sum_k h_k \left(\frac{\partial}{\partial x}\left(\rho D_k(T)\frac{\partial Y_k}{\partial x}\right)\right) + \rho \sum_k h_k \left(\frac{\partial Y_i}{\partial t}\right)_{ch} - \rho_p u_p \frac{(u-u_p)}{\tau_{St}} - \rho_p c_{P,p} Q, \tag{4}$$

$$P = R_B T n = \left(\sum_i \frac{R_B}{m_i} Y_i\right)\rho T = \rho T \sum_i R_i Y_i, \tag{5}$$

$$\varepsilon = c_v T + \sum_k \frac{h_k \rho_k}{\rho} = c_v T + \sum_k h_k Y_k, \tag{6}$$

$$\sigma_{xx} = \frac{4}{3}\mu\left(\frac{\partial u}{\partial x}\right) \tag{7}$$

We use here the standard notations: $P$, $\rho$, $u$, are pressure, mass density, and flow velocity of gaseous mixture, $Y_i = \rho_i/\rho$ - the mass fractions of the species, $E = \varepsilon + u^2/2$ - the total energy density, $\varepsilon$ - the internal energy density, $R_B$ - is the universal gas constant, $m_i$ - the molar mass of i-species, $R_i = R_B/m_i$, $n$ - the gaseous molar density, $\sigma_{ij}$ - the viscous stress tensor, $c_v = \sum_i c_{vi} Y_i$ - is the constant volume specific heat, $c_{vi}$ - the constant volume specific heat of i-species, $h_i$ - the enthalpy of formation of i-species, $\kappa(T)$ and $\mu(T)$ are the coefficients of thermal conductivity and viscosity, $D_i(T)$ - is the diffusion coefficients of i-species, $(\partial Y_i / \partial t)_{ch}$ - is the variation of i-species concentration (mass fraction) in chemical reactions, $\rho_p = m_p N_p -$ mass density of the suspended particles, $N_p$ – particles number density, $u_p$ – particles velocity, $\tau_{St} = m_p / 6\pi\mu r_p$ - the Stokes time for the spherical particles of the radius $r_p$ and mass $m_p$, $Q$ – interphase thermal exchange source, $c_{P,p}$ – the constant pressure specific heat of the particles material.

The changes in concentrations of the mixture components $Y_i$ due to the chemical reactions are defined by the solution of system of chemical kinetics

$$\frac{dY_i}{dt} = F_i(Y_1, Y_2, ... Y_N, T), \quad i = 1, 2, ... N. \tag{8}$$

The right hand parts of (8) contain the rates of chemical reactions for the reactive species $H_2$, $O_2$, $H$, $O$, $OH$, $H_2O$, $H_2O_2$ and $HO_2$ with subsequent chain branching, production of radicals and energy release. We use in simulations the standard reduced chemical kinetic scheme for hydrogen/oxygen combustion with the elementary reactions of the Arrhenius type and with pre-exponential constants and activation energies presented in [26]. This reaction scheme for a stoichiometric $H_2/O_2$ mixture has been tested in many applications and reproduces the main features of the chain branching reaction to a large extent adequate to complete chemical kinetics. The computed thermodynamic, chemical, and material parameters using this chemical scheme are in a good agreement with the flame and detonation parameters measured experimentally. For example, for $P_0 = 1.0$ bar we obtain for the laminar flame velocity, the flame thickness and adiabatic flame temperature $U_f \approx 12 \text{m/s}$, $L_f = 0.24 \text{mm}$, $T_b = 3100 \text{K}$, correspondingly, for the expansion ratio (the ratio of the density of unburned gas and the combustion products) $\Theta = \rho_u / \rho_b = 8.3$, and temperature and velocity of CJ-detonation $T_{CJ} = 3590 \text{K}$, $U_{CJ} = 2815 \text{m/s}$.

The equations of state for the reactive mixture and for the combustion products were taken with the temperature dependence of the specific heats and enthalpies of each species borrowed from the JANAF tables and interpolated by the fifth-order polynomials [27]. The viscosity and thermal conductivity coefficients of the mixture were calculated from the gas kinetic theory using the Lennard-Jones potential [28].

$$\mu = \frac{1}{2}\left[\sum_i \alpha_i \mu_i + \left(\sum_i \frac{\alpha_i}{\mu_i}\right)^{-1}\right], \qquad (9)$$

where $\alpha_i = \frac{n_i}{n}$ is the molar fraction, $\mu_i = \frac{5}{16}\frac{\sqrt{\pi \hat{m}_i kT}}{\pi \Sigma_i^2 \tilde{\Omega}_i^{(2,2)}}$ is the viscosity coefficient of $i$-species, $\tilde{\Omega}^{(2,2)}$ - is the collision integral which is calculated using the Lennard-Jones potential [26], $\hat{m}_i$ is the molecule mass of the i-th species of the mixture, $\Sigma_i$ is the effective molecule size. The thermal conductivity coefficient of the mixture is

$$\kappa = \frac{1}{2}\left[\sum_i \alpha_i \kappa_i + \left(\sum_i \frac{\alpha_i}{\kappa_i}\right)^{-1}\right]. \qquad (10)$$

Coefficient of the heat conduction of i-th species $\kappa_i = \mu_i c_{pi}/\text{Pr}$ can be expressed via the kinematic viscosity $\mu_i$ and the Prandtl number, which is taken $\text{Pr} \approx 0.75$.

The binary coefficients of diffusion are

$$D_{ij} = \frac{3}{8}\frac{\sqrt{2\pi kT \hat{m}_i \hat{m}_j / (\hat{m}_i + \hat{m}_j)}}{\pi \Sigma_{ij}^2 \tilde{\Omega}^{(1,1)}(T_{ij}^*)} \cdot \frac{1}{\rho}, \qquad (11)$$

where $\Sigma_{ij} = 0,5(\Sigma_i + \Sigma_j)$, $T_{ij}^* = kT/\varepsilon_{ij}^*$, $\varepsilon_{ij}^* = \sqrt{\varepsilon_i^* \varepsilon_j^*}$; $\varepsilon^*$ are the constants in the expression of the Lennard-Jones potential, and $\tilde{\Omega}_{ij}^{(1,1)}$ is the collision integral similar to $\tilde{\Omega}^{(2,2)}$ [28]. The diffusion coefficient of i-th species is

$$D_i = (1 - Y_i) / \sum_{i \neq j} \alpha_i / D_{ij} . \tag{12}$$

A detailed description of the equations and transport coefficients used for the gaseous phase and calculation of diffusion coefficients for intermediates has been published previously [29-32].

The equations for solid particles is taken in continuous hydrodynamic approximation. The interaction between particles is negligibly small for a small volumetric concentration of particles, so that only the Stokes force between the particle and gaseous phase is taken into account. Then the equations for the phase of suspended particles read:

$$\frac{\partial N_p}{\partial t} + \frac{\partial (N_p u_p)}{\partial x} = 0 \tag{13}$$

$$\left( \frac{\partial u_p}{\partial t} + u_p \frac{\partial u_p}{\partial x} \right) = \frac{(u - u_p)}{\tau_{St}}, \tag{14}$$

$$\left( \frac{\partial T_p}{\partial t} + u_p \frac{\partial T_p}{\partial x} \right) = Q - \frac{2\pi r_p^2 N_p}{c_{P,p} \rho_{p0}} \left( 4\sigma T_p^4 - q_{rad} \right). \tag{15}$$

Here $T_p$ – temperature of the particles, $2\pi r_p^2 N_p \left( 4\sigma T_p^4 - q_{rad} \right)$ is the thermal radiation energy accumulated and re-emitted by the particles. The energy transferred from the particle surface to surrounding gaseous mixture is

$$Q = (T_p - T) / \tau_{pg}, \tag{16}$$

where $\tau_{pg} = 2 r_p^2 c_{P,p} \rho_{p0} / 3 \kappa Nu$ is the characteristic time of the energy transfer from the particle surface to surrounding gaseous mixture, $c_{P,p}$ and $\rho_{p0}$ are specific heat and mass density of the particles, Nu is the Nusselt number (see e.g. [33]).

For a planar problem the equation for the thermal radiation heat transfer in the diffusion approximation is [34, 35]:

$$\frac{d}{dx}\left(\frac{1}{\chi}\frac{dq_{rad}}{dx}\right) = -3\chi\left(4\sigma T_p^4 - q_{rad}\right), \tag{17}$$

where the radiation absorption coefficient is $\chi = 1/L = \pi r_p^2 N_p$, and $L = 1/\pi r_p^2 N_p$ is the radiation absorption length. The radiation energy flux emitted by the hot combustion products from the flame front is assumed to be equal to the blackbody radiative heat source, $q_{rad}(x = X_f) = \sigma T_b^4$, where $\sigma = 5.6703 \ 10^{-8} \text{W}/\text{m}^2\text{K}^4$ is the Stefan-Boltzmann constant, and $T_b \approx 3000\text{K}$ is temperature of the hot combustion products.

The calculations were carried out for stoichiometric hydrogen/oxygen at initial pressure $P_0 = 1\text{atm}$ with a small solid inert spherical particles suspended in the gaseous mixture. For simplicity the particles are assumed to be identical and the mass density of the particles, $\rho_p = m_p N_p$, is taken much smaller than the gas density, $\zeta = m_p N_p / \rho \ll 1$, so that there is only one way of a momentum coupling of the particles and the gaseous phase. The numerical method is based on splitting of the Eulerian and Lagrangian stages, known as coarse particle method (CPM) [36]. Detailed description of the modified CPM scheme, together with the convergence and resolution tests are presented in Appendixes A and B.

## 3. Flame acceleration in the mixture with uniformly dispersed particles

In this section we consider the flame propagating through the particle-laden two phase mixture with uniformly suspended particles. As it was mentioned above, the particles ahead of the flame front absorb the thermal radiation and then lost the heat by conduction to the surrounding unreacted gaseous phase. In case of the uniformly suspended particles it causes the preheating of the gaseous mixture ahead of the flame on the scales of the thermal radiation absorption length (see Fig. 1). The calculated profiles of the gaseous phase and the particles temperature, which are established for the radiation absorption lengths, L=0.28, 1.11, and 2.22cm, after the stationary flow was settled, are shown in Fig.2. The radiation absorption lengths, L=0.28, 1.11, and 2.22cm correspond to three variants: $r_p = 6.0; 1.5; 0.75 \mu m$ and $N_p = 3.18 \cdot 10^6; 1.27 \cdot 10^7; 2.55 \cdot 10^7 cm^{-3}$. The characteristic time scales of the problem for these cases are: $\tau_{St} = 170 \mu s$, $\tau_Q \approx 130 \mu s$; $\tau_{St} \approx 10 \mu s$, $\tau_Q \approx 8 \mu s$; and $\tau_{St} \approx 2.6 \mu s$, $\tau_Q \approx 2 \mu s$, correspondingly. The maximum temperature of the particles ahead of the flame and the maximum increase of the flame velocity are established during the characteristic hydrodynamic time of the order of $t \approx L/U_{f0}$, where $U_{f0}$ is the normal laminar flame velocity in pure gas mixture. It should be noticed that if the characteristic time of the energy transfer from the particle surface to surrounding gaseous mixture, $\tau_Q$, is smaller than the characteristic gasdynamic time of flame propagation $L_f / U_f \approx 20 \mu s$, then thermal exchange between the gaseous phase and the particles is fast, and the particles and gas temperatures are not noticeably different. A momentum coupling of the particles and the gaseous phase is essential if the mass loading parameter (see e.g. [37]) $\varsigma = m_p N_p / \rho$ becomes of the order of unity. In order to distinguish the effects of the radiation preheating in all calculations the mass loading parameter was taken the same, $\varsigma = 0.3$, for all the calculated variants. In this case the initial temperatures and the flame velocities are the same, so that this allows to distinguish the impact of the radiation preheating on the flame dynamics.

Absorption of the radiation by the particles and the temperature increase of the reacting gaseous mixture ahead of the flame result in the corresponding increase of the flame velocity. As shorter the absorption length is as shorter the preheated zone ahead of the flame, and as faster the flame velocity increases up to some new constant value. The corresponding increase of the combustion wave velocities due to the temperature increase ahead of the flame for the radiation absorption lengths L=0.28, 1.11, 2.22 cm is shown in Fig. 3. The increase in the combustion velocity is modest, about 10%, and it is maximal for the given parameters. Taken into account that the stationary flow is established within the time of the order of $L/U_f$, it is straightforward to estimate the maximum temperature increase ahead of the flame due to the radiative heat transfer in the laden-particle mixture. In the coordinate system co-moving with the flame front the mixture with suspended particles flows towards the flame with the velocity $U_f$. The thermal radiation is noticeably absorbed by the particles which are closer than L to the flame front. Thus, the characteristic time of the radiation heating for the Lagrangian particle initially distant at L before it arrives to the flame front is $t = L/U_f$. Using the energy balance equations we obtain for the maximum increase of temperature of the unburned mixture close to the flame the following expression

$$\Delta T = \sigma T_b^4 \frac{1}{U_f} \frac{(1-e^{-1})}{(\rho_p c_p + \rho c_{V,g})} . \qquad (18)$$

As it can be readily observed from (18) the maximum temperature increase due to the radiation preheating does not depend on the radiation absorption length. This is due to the fact that although the local radiative heating for smaller absorption length is larger, but for larger absorption lengths the particles absorb the radiant heat flux over a longer time until the stationary flow is established, and this compensates their less local heating. The adiabatic flame temperature is less than it is in a pure mixture, which is diluted in the presence of neutral particles. The photons emitted from the flame front are produced within the radiative layer near the flame front of the finite thickness, so that the effective temperature of the radiating flame surface is also lower. Taking the effective temperature for the radiation emitted from the flame surface $T_{b,eff} \approx 2700K$ the maximum temperature increase caused by the radiation preheating according to the estimate (18) is $\Delta T \approx 160K$ that in a good agreement with the values obtained in the numerical simulations in Fig. 2.

The radiation heating time and the attained during the radiation preheating maximum temperature ahead of the flame as greater as smaller the normal laminar flame velocity in a pure gaseous mixture. As smaller the laminar flame velocity in a pure gaseous mixture as longer is time of the radiation preheating and according to Eq. (18) as higher is the maximal temperature of the particles and the mixture attained ahead of the flame.

At sufficiently low initial pressure or for the flame with essentially small normal laminar velocity the maximum temperature ahead of the flame caused by the radiation preheating can exceed the crossover temperature, which determines the limit of the induction time where the induction endothermic stage passes into the fast exothermic stage. In this case the thermal radiation heat transfer in the particle-laden combustible mixture becomes a dominant process and the mechanism of combustion wave propagation is the sequence of independent ignitions in the region of radiation absorption length ahead of the flame front instead of usual gaseous thermal conduction. It is straightforward to obtain an order-of-magnitude estimate for the speed and width of the flame front using a dimensional analysis similar to how this is done for normal laminar flame [38]. In case when the temperature ahead of the flame caused by the radiation preheating exceeds the crossover value an order-of-magnitude of the flame width is $L_{fR} \sim L$, and the flame velocity can be estimated as $U_{f,rad} \propto (L/\tau_R) = (L/L_f)U_{f0} \gg U_{f0}$. The velocity $U_{f,rad}$ of such flame exceeds considerably the laminar flame velocity in a pure gaseous mixture by large factor $(L/L_f) \gg 1$. Such a combustion wave will look like a sequence of thermal explosions, accompanied by a strong increase in pressure. Presumably explosions caused by accidental ignition of natural gas and coal dust accumulations in underground coal mines are not necessarily the onset of detonation, but could be associated with the transition of conventional combustion mode to the above described combustion regime in coal dust clouds, when the mechanism of combustion wave propagation is dominated by the radiant heat transfer and its absorption by the suspended particles of coal dust.

Presumably, the radiation heat transfer in particle-laden gaseous mixture, when the thermal radiation heat transfer becomes a dominant mechanism of the flame propagation, is inherent to the particle-laden flames only, and is a unique situation when the thermal radiation heat transfer can be a dominant mechanism of the flame propagation. In this case there is no thermal equilibrium between the gaseous phase and the radiation. The radiant heating of the mixture occurs through two stages: 1) solid particles absorb the radiant flux; 2) heated particles transfer the heat to the surrounding gas. Interesting that even in the case of a thermonuclear combustion when radiation can be in thermal equilibrium with the surrounding medium it is unlikely that the radiation heat transfer can be a dominant process. Assuming for the sake of simplicity a black body thermal radiation, taking for the radiation heat transfer $q_{rad} = (16/3)\sigma L T^3 dT/dx$, it is readily seen that the coefficient of electron thermal conduction $\kappa_e = (1.3/\Lambda Z) \cdot 10^{11} T_e^{5/2}$ exceeds the coefficient of radiation heat transfer $\kappa_{rad} = 16\sigma L T^3/3$ even at thermonuclear temperatures, $T = (5 \div 10)\,\text{keV}$. Only in thermonuclear combustion of strongly Fermi degenerated matter of White dwarf the radiation heat transfer became comparable but still smaller than the electron thermal conduction $\kappa_e \propto 10^{10}\,T/K$ [39].

For the radiation heat transfer in particle-laden gaseous mixture to become a dominant mechanism of the flame propagation the radiation heating must be relatively long to rise temperature ahead of the flame up to the crossover value before the arrival of the original advancing flame. In principle it is possible either for a flame in the gas mixture of low density, or in a low reactive mixture with a small enough laminar flame velocity. As a result of the radiation preheating, an inhomogeneous in space temperature distribution can be formed in the unburned mixture ahead of the flame with the steepness of the temperature gradient determined by the thermal radiation absorption length. If the maximum temperature within the formed temperature gradient exceeded the crossover value, then depending on the steepness of the temperature gradient either deflagration or detonation can be ignited via the Zeldovich's gradient mechanism [25]. The possibility of this depends on the competition between the rate of ignition and the characteristic hydrodynamic time: the exothermal stage of chemical reaction must develop before the initial flame arrival. Since the minimum scales of the temperature gradient initiating various modes of combustion increase significantly with the decrease of initial pressure and for slow-reacting mixtures [40, 41] such modeling is very much time consuming. For example, the minimum scale of the linear temperature gradient capable to trigger detonation in the hydrogen oxygen mixture is approximately 10cm at the initial pressure 1atm and it is greater than one meter at the initial pressure 0.1atm and for the hydrogen air mixture at initial pressure 1atm. The numerical simulation of the problem in question requires fine resolution for a pattern size of several meters and it is time consuming even for the 1D problem. It is unlikely that detonation can be ignited via the Zeldovich's gradient mechanism in such slow reactive mixture as the methane-air, as the minimum scale of the temperature gradient in this case is extremely large (a very shallow gradient is required) that is practically senseless. Instead, in the next section we will demonstrate ignition of different combustion modes by the temperature gradient formed via the radiative preheating in the hydrogen oxygen particle-laden mixture for the case of non-uniform spatial distribution of micro particles.

## 4. Combustion regimes initiated by the radiative preheating of the mixture with non-uniformly dispersed particles

Since the radiant energy flux emitted from the flame is not very intense, temperature of the gaseous mixture ahead of the flame does not increase significantly for a uniform distribution of the particles. It results in a relatively small increase of the burning velocity and causes modest effect on the overall process evolution. The interesting opportunities appears in case of non-uniform distribution of the particles (layered dust cloud), such that over a noticeably large distance ahead of the flame the mixture is almost transparent for the thermal radiation and the radiation is absorbed only far ahead of the flame front as it is shown schematically in Fig. 4. It is a well established phenomenon that in coal mines the pressure wave of a weak accidental methane explosion can disperse dust deposits leading to the formation of a non-uniformly suspended particles forming a layered dust-air cloud. In this case time of the thermal radiation heating is much longer than in the case of a uniform particles distribution as the flame should overcome the ultimate distance separating it from the dense region. Particles in the dense cloud far ahead from the advancing flame absorb much more of the thermal radiant energy before the flame arrival to the boundary of denser cloud, so that temperature of the particles and surrounding mixture can rise up to the value suitable for ignition long before the flame arrival. In such conditions the ignition starts when the temperature exceeds the crossover value, which is for hydrogen/oxygen at 1atm is about 1050±50K and determines the limit where the induction endothermic stage passes into the fast exothermic stage. An order-of-magnitude estimate of the time interval needed to rise temperature of the suspended particles up to the crossover temperature is about 1ms . This means that for absorption length inside the cloud L>1cm, the left boundary of the denser particles cloud, where radiation is absorbed, should be well ahead from the propagating flame as shown in Fig. 4. Below we assume that the "gap" between the flame and the left boundary of the gas/particles cloud is transparent for the thermal radiation, so that in this area the radiation absorption length is much larger than the distance from the flame to the boundary of the cloud.

The temperature gradient established due to the radiative preheating of the mixture in the gas-particle cloud via radiation absorbed by the particles depends on the thermal radiation absorption length, which is also influenced by the gas expansion during the heating. If the particle-gas cloud is far enough from the flame so that the combustible gas temperature rises up to crossover value, then, depending on the steepness of the formed temperature gradient, either a deflagration or a detonation can be ignited via the Zeldovich's gradient mechanism [40-41].

Figure 5 shows the calculated time evolution of the gaseous temperature (5a) and mass density of the particles profiles (5b) during preheating for the distant cloud with the initial stepwise profile of particles number density. In the calculations the particle radius and the particles number density were taken: $r_p = 1\mu m$ and $N_p = 2.5 \cdot 10^7 cm^{-3}$, correspondingly. The number density of particles between the left boundary of the cloud and the flame is assumed to be much less than it is inside the cloud, so that the radiation absorption length there is much larger than the distance between the flame and the left boundary of the cloud and this "gap" can be treated as fully transparent for the radiation. At the same time it should be noticed that even relatively small concentration of the particles in the hot combustion products behind the flame front increases considerably the luminosity and the radiant energy flux, which justifies the black-body radiation flux emitted from the flame surface. The steepness of the formed temperature gradient in Fig. 5, $\Delta = (T^* - T_0)/|dT/dx| \approx 1cm$, is in agreement with the value of the radiation absorption length $L = 1/\pi r_p^2 N_p \approx 1.2 cm$. Such temperature gradient can ignite a deflagration according to the classification of combustion regimes initiated via the Zeldovich's gradient mechanism in the hydrogen/oxygen at 1 atm [40, 41]. The calculated time evolution of the gaseous temperature and pressure profiles during the deflagration wave formation in the vicinity of the left boundary of the gas-particles cloud far ahead of the flame front are shown in Fig. 6. Here the dashed line in Fig. 6(a) shows the initial density profile of the particles and the solid line shows the formed density profile at the instant $t_0$=890μs prior to the ignition, when the corresponding temperature gradient with $T^* = 1050\,K$ shown in the first temperature profile in Fig.6(b), is formed.

For the case of a somewhat smaller initial number density of the particles in the cloud layer ahead of the advancing flame the temperature gradient, formed at the moment, when maximal temperature $T^*$ achieved crossover value, is shallower and the fast deflagration behind a weak shock wave can be formed [40, 41]. It should be noticed that the principle factor which defines steepness of the temperature gradient caused by the radiative preheating is the radiation absorption length, $L = 1/\sigma_p N_p$. A more gentle temperature gradient can be formed either in the cloud with smaller concentration of the particles, or for smaller particles of the same concentration, or for the cloud with a properly diffuse interface instead of the cloud with the stepwise number density spatial distribution of the particles. In the latter case the radiation absorption length varies along the diffusive cloud interface, which may result in the formation of a smoother temperature profile with a shallower temperature gradient capable to initiate a detonation. Examples of the particle clouds with diffuse interface and the corresponding calculated temperature profiles caused by the radiative preheating are shown in Figs.7 and 8.

We consider the cloud of particles with the same maximal number density of particles and with a smooth, diffuse left boundary. The upper frame in Figure 7(a) shows the initial (dashed line) and the formed diffuse boundary of the particles cloud (solid line) and time evolution of the gaseous temperature and pressure profiles during the formation of the fast deflagration behind the outrunning weak shock in the vicinity of the gas-particles cloud boundary. The initial boundary diffusivity of the cloud is the linear decrease of the particles number density on the scale of 1cm.

Initiation of a detonation wave via the Zeldovich's gradient mechanism requires much more shallower temperature gradient. Calculations [40, 41] for hydrogen/oxygen, which used the initial linear temperature gradient at normal conditions and $T^* = 1050\,K$ at the top of the gradient, have shown that the scale of the gradient needed for the detonation initiation was about 20cm. The results of the simulation for the cloud with initial diffuse boundary, where the particles number density drops linearly on the scale of 10cm, are presented in Figure 8. Here the upper frame, Fig. 8(a), depicts the formed diffuse boundary of the particles cloud, which was smeared due to gas expansion during the thermal radiation heating up to 20cm. Time evolution of the gaseous temperature and pressure profiles during the detonation formation in the vicinity of the diffusive cloud boundary are depicted in Fig.8(b). Dashed line in the upper frame shows the initial density profile, and the solid line shows density profile at the time instant $t_0 = 5015\mu s$ prior to the ignition of a spontaneous combustion wave triggering the detonation. Figure 9 shows the evolution of the spontaneous reaction wave speed in the laboratory referenced frame for the three scenarios presented in Figs. 6, 7, 8. It can be noted that the fast flame speed decreases as the shock outruns and it saturates at almost the same value as the slow deflagration (1) propagating inside the dense particle cloud and accelerated due to the radiant preheating mechanism considered in section 3. Thus, the mechanism of radiant preheating is the principle one for the flame propagation inside the dense particle clouds. The radiant preheating does not influence supersonic detonation wave, and it propagates with almost the Chapman-Jouguet velocity with almost negligible momentum and heat losses because of relatively small particles number density.

**Discussion and Concluding remarks**

The purpose of the present study was to demonstrate that effects of the radiative heat transfer can considerably influence the overall picture of the propagation of planar particle-laden dust flames. Depending on the spatial distribution of the particles, the radiative preheating can either result in the increase of the flame velocity in the case of uniform distribution of the particles, or in the case of non-uniform distribution of the particles it can promote formation of the temperature gradients inside the distant dense gas-particles cloud, which can ignite either deflagration or detonation via the Zeldovich's gradient mechanism far ahead of the advancing flame. The performed numerical simulations demonstrate the plausibility of radiation preheating as the principal effect of the combustion intensification and even detonation initiation in the gaseous fuel, where relatively low concentration of suspended solid particles or any other substance can absorb the radiative heat flux and rise temperature of the fuel ahead of the flame. It should be emphasized that this study is a necessary prerequisite aiming to show principle physics and role of the radiative preheating. The obtained results show that the thermal radiative preheating can play a significant role in determining the regimes of combustion in two-phase reacting flows.

It must be emphasized that the effects of the radiation heat transfer is much stronger for a slow flame, such as methane-air flame in the coal dust cloud. In the later case the radiative preheating for a uniformly dispersed particles can rise temperature ahead of the advancing flame up to 900K, so that the flame velocity increases in 5-6 times. It is well known that many experiments have shown that methane-air could sustain a detonation only if it was initiated by very strong external ignition source producing a strong shock wave, however experiments at least conducted within smooth tubes were not able to achieve detonation via DDT. Since the explosions in underground coal mines begin from an accidental flame ignited usually by e.g. a small spark, it is unlikely that the deflagration-to-detonation transition can be origin of such explosions.

Possible new scenarios of origin of the explosions in underground coal mines related to the effects of the radiation heat transfer in a layered dust-air clouds were for the first time proposed by Liberman [42]. In coal mines the pressure wave of a weak methane explosion or convective flows from the external sources can disperse dust deposits leading to the formation of a layered dust-air cloud. A layered coal dust-air clouds are a well known phenomenon in coal mines. The case of the layered dust cloud is characterized by relatively longer length and time scales of the radiation preheating similar to that shown in Figs. 4-9 for hydrogen-oxygen mixture. It is straightforward to obtain that for the radiation emitted by the advancing flame time needed to rise temperature in the nearest denser layer up to the value $\Delta T \approx (1500 \div 1800)K$ is $\delta t = (\rho_p c_p + \rho c_{V,g}) \Delta T / \sigma T_b^4 \approx 1ms$. The induction time of methane-air for such temperatures is less than 1ms. This means that the radiation of the primary flame causes ignition of a new flame in the nearest denser layer. The radiation of the secondary flame causes ignition of the flame in the next denser layer, etc. As a result of such successive ignition, combustion will propagate with the velocity which magnitude is defined by the spacing between the coal dust layers and can be as high as 1000m/s. This scenario of successive ignitions, which looks like a volume thermal explosions can be accompanied by a noticeable pressure increase and it is likely explain origin of the explosions in underground coal mines.

In view of the results presented in this paper there can be interesting possibility explaining scenario of thermonuclear burning of type Ia Supernovae. SNe Ia are invaluable for measuring the large-scale structure of the Universe and for the determination of the Hubble constant. The peak brightness of SN Ia light curves is correlated with its duration: more luminous supernovae display the reduced rate of the peak brightness which is related to the amount of $^{56}Ni$ formed during the explosion. It is possible to normalize all SNe Ia to the same peak luminosity according to their light curves, which is the basis for light curve calibration that allows use them as unique standard candles. However, even today the measured value of the Hubble constant is uncertain to about 10%. For a correct theoretical description of SN Ia light curves it is necessary to model correctly SN Ia explosion.

In the context of thermonuclear burning of type Ia Supernovae, combustion initially proceeds in the deflagration mode with the velocity and width of a laminar flame at the central densities $\rho_c = 2 \cdot 10^9 \, g/cm^3$ being $U_f \approx 10^7 \, cm/s$ and $10^{-2} \, cm$. Type Ia supernovae (SNe Ia) begins with a white dwarf (WD) near the Chandrasekhar mass that ignites a degenerate thermonuclear runaway close to its center and explodes having initial radius $R_{WD} \approx 10^8 \, cm$. Such a wide range of the length scales necessitates to use models of the infinitely thin flame (see [43] for a review of explosion scenarios) that limits sufficiently the understanding of the transient phenomena.

Nowadays the best modeling for majority of the observed events is provided by the so-called delayed detonation models, which imply a phase of subsonic thermonuclear burning (deflagration) during which the star expand and a phase of a detonation, which burns remaining fuel on timescales much shorter than the timescale of the explosion. The paradigm of the delayed detonation models is consistent with the theoretical predictions [44] that the detonation in a strongly degenerate matter is unstable against 1D pulsations at densities higher than $2 \cdot 10^7 \, \text{g/cm}^3$ and it becomes stable at densities smaller than $2 \cdot 10^7 \, \text{g/cm}^3$ near the star surface. There has been numerous attempts both analytical and numerical [45-50] to explain the detonation formation in SN Ia explosion. However, after many years of studies a fundamental question what is the mechanism of deflagration-to-detonation transition (DDT) in a white dwarf star is still unclear (see e.g. [50]).

We want to draw attention to possible new scenario, which can explain DDT in SN Ia explosion. During the late phase of the deflagration regime the radiant flux produced by the radioactive decays of $Ni^{56} \to Co^{56} \to Fe^{56}$ increases considerably in course of the star incineration by expanding deflagration. Absorption of the radiant energy flux in the outer layers of the star can produce a sufficient preconditioned regions such that a detonation arises via the Zel'dovich mechanism.

**Acknowledgements**

We acknowledge the allocation of computing resources provided by the Swedish National Allocations Committee at the Center for Parallel Computers at the Royal Institute of Technology in Stockholm, the National Supercomputer Centers in Linkoping and the Nordic Supercomputer Center in Reykjavik. This work was carried out within the program supported by Swedish Research Council and the Research Council of Norway under the FRINATEK, Grant 231444 (M.L.). The Ben-Gurion University Fellowship for senior visiting scientists is acknowledged (M.L.). Partly the work was supported by the Program of Russian Academy of Sciences (MI, AK). The results of this study have been presented at theoretical and astrophysics seminars at Ben-Gurion University and NORDITA workshops and conferences and M.L. acknowledges useful discussions with N. Kleeorin, A. Brandenburg, N. Haugen and I.Rogachevskii during writing this paper.

**Appendix A: The numerical method**

Numerical simulations of the reacting flows deal with the time-dependent Navier-Stokes equations which include strongly nonlinear terms in the right hand part and become stiff if the characteristic time scales of the corresponding processes (chemical kinetics, relaxation etc.) are much less than characteristic gasdynamical time scales [51-56]. The stiff terms in the equations demand using specific methods to achieve reliable solution. There are two basic options to tackle and to overcome difficulties related with the stiff problems (see e.g. [51]). We either can use the first-order schemes on the grids with very fine computational cells or we can use high-order schemes with limiters. In the case if, for example, we need a solution of the large-scale combustion problems, where the thickness of the reaction zone is by many orders of magnitude less than the characteristic scales of the problem, the use of the first order schemes would require too much computational recourses since the use of rather coarse grids does not provide enough resolution to suppress numerical viscosity for a reliable solution. On the other hand, the approach using high order schemes requires development and utilization of the specific limiters necessary to suppress unphysical oscillations intrinsic to the high orders schemes (higher then the second one). Often such limiters are developed for a certain classes of problems [51-56] that make them inapplicable for solving a wider class of problems.

In many cases it is advantageous to use a simple numerical scheme of the first or second order instead of more complex methods of the high order schemes as one should not analyze in details and chose appropriate individual measures to suppress the unphysical oscillations. For the one-dimensional problems studied in the present work the main task was to resolve physical effects on relatively small spatial scales. For that purpose we used the second order in space numerical scheme taking numerical grids with as fine as possible cells to resolve the reaction zones in details at the background of a large-scale flow. In this case there are no unphysical oscillations in the solution and the main limitation determining the accuracy of the solution is a fine resolution of the contact surfaces smoothed over the grid due to relatively high scheme viscosity. Here this problem was overcame by means of usage of high resolution grids which also is dictated by the necessity to resolve structure of the flame front. Contrary to [54-56], where the detonation problem was solving, in case of studying transient evolution of the ignition and combustion processes one should resolve the flame front thickness, as just the flame front determines the principal spatial scale until the onset of a steady detonation. Below we give a short description of the used numerical method and in the Appendix B we present resolution tests which show convergence of the obtained solutions.

The numerical method used in the present study is the second order numerical scheme based on splitting of the Eulerian and Lagrangian stages, also known as the coarse particle method (CPM). For the first time such scheme was introduced by Gentry, Martin, and Daly [57] and afterwards it was modified and widely implemented by Belotsercovskii and Davydov [36, 58, 59]. This scheme appears to be rather robust when used to model various kinds of complex hydrodynamic flows. The analysis of this method applied to a large variety of problems, including problems of hydrodynamic instability, have shown that it possesses a high numerical stability, which enables to carry out calculations of shock waves without the aid of artificial viscosity. High stability of the method is achieved by dividing one time-step calculation into three stages. On the first stage, the change of hydrodynamic characteristics on the fixed Eulerian space grid is calculated using the explicit scheme without regarding of mass, momentum and energy transfer. The hydrodynamic variables are transferred through the cell boundaries on the second stage using the values of hydrodynamic characteristics from the first stage. The last, third, stage consists in final calculation of the values of all the parameters for every cell and for the whole system. Thermal conductivity and diffusion are calculated on the third stage but energy is calculated partially on the first stage and partially on the third one.

The system of chemical kinetics equations represents a stiff system of differential equations, and it was solved using Gear's method [60]. The developed algorithm was implemented using the FORTRAN-90. The simulations were performed using high-performance parallel super-computer. The choice of an optimal approximation scheme for stress tensor and for the fluxes through the cell boundaries enabled to achieve high stability in the calculation of the flame dynamics. It was shown that a high numerical stability of the method is achieved if the hydrodynamic variables are transferred across the grid boundary with the velocity, which is an average value of the velocities in neighboring grids. The overall modified solver is of the second-order accurate, providing high accuracy of the solutions. The modified CPM and optimal approximation scheme were thoroughly tested and successfully used for simulation engine combustion knock occurrence in SI engine [61, 62] and other problems in the field of transient combustion and detonation phenomena [29-32, 40, 41].

**Appendix B: Code validation, resolution and convergence tests**

While solving the problem of laminar flame propagating through the gaseous mixture uniformly seeded with the particles considered in section 3 one should use a proper resolution that allows exclude the errors and artificial solutions. Therefore the thorough convergence and resolution tests were carried out to verify that the observed phenomena were correctly caught remaining unchanged with increasing resolution.

Figure A1 represents the results of the convergence test for a combustion wave propagating through the non-seeded stoichiometric hydrogen-oxygen mixture at normal conditions ($T_0 = 300K$, $p_0 = 1 atm$ p0=1atm). Figure A2 shows similar test but for the flame propagating in the mixture seeded with the particles (case $L = 2.22 cm$) considered in section 3. In the latter case the combustion velocity and the flame structure is fully defined by the state of the gaseous mixture just ahead of the flame front. As the temperature in this region does not exceed (450÷500)K and the pressure remains almost constant the convergence test gives almost the same fine resolution as in the case of non-seeded mixture. As it was mentioned in Sec.3 in case of considerably smaller combustion velocities the radiative preheating can be more efficient close to the flame front. In such conditions one should resolve flame structure at the higher temperatures and use finer meshes, as it is seen in Fig. A3, which shows resolution tests for different ambient temperatures with grey signs for acceptable range of convergence. The meshes were taken to resolve the structure of the flame front with 6, 8, 12, 24, and 48 computational cells, corresponding to the computational cell sizes: $\Delta$ = 0.1, 0.05, 0.025, 0.01 and 0.005mm, respectively. The acceptable quantitative convergence was found for resolution of 24 computational cells per flame front (see Figs. A1, A2). Therefore this or higher resolution was typically used for solving the problems in Section 3.

The problems considered in Section 4 demands more robust parameters of the computational setup. The most demanding is the case where the detonation arises as a result of auto-ignition inside the hot-spot (see Fig. 8). Let first formulate the consequence of the processes taking place in such a case that should be appropriately resolved. As it was shown in [30] the detonation caused by the Zel'dovich gradient mechanism arose via the following scenario: 1) the spontaneous combustion wave were formed, 2) as the spontaneous wave decelerated the pressure wave formed behind its front, overran it forming the shock wave, 3) detonation established after the transient process involving flame acceleration in the flow behind the outgoing shock wave.

Due to this sequence of the events one should appropriately resolve the combustion waves propagating through reacting medium at initially elevated temperature (~1000K, see figures in Section 4) and on the background of elevated temperature and pressure behind the shock front. Besides the coupling of the reaction wave and shock should be resolved taking into account that the flame thickness is much larger than the width of the shock front. According to this we performed a number of resolution tests for the flames propagating at elevated temperatures and pressures (see e.g. Fig. A3) and also in Figure A4 present the resolution test for the problem considered in Fig. 8. Figure A5 shows the burning velocity-temperature dependence. One can observe that agreement between the calculated data and the extrapolation of the experimental data [63] using the global reaction order $n = 2.74$ is satisfactory good. The error bars for calculations are due to the lack of information about the ignition conditions. The ignition conditions in the calculations were such that the compression wave arose and outran from the ignition zone preheating the mixture ahead the forming flame front that could cause the discrepancy with the experimental data [63]. As it is seen from Fig. A3, which shows the rates of convergence at different resolutions for three different initial temperatures, the flame dynamics at the elevated temperatures can be resolved only with a finer resolution than at the lower temperatures. Therefore to obtain the converged solution for the detonation initiation problem one should use a finer resolution from the very beginning. For the model considered in the present paper the resolution was taken with at least 48 computational cells per flame front at normal conditions, that agrees well with the results obtained previously in [30]. Figure A4 shows the convergence tests for the characteristic time scales determining the stage of radiant preheating ($t_0$) and transient stage of detonation formation ($\tau_D$) in case presented in Fig. 8. The results are almost the same as in Fig. A3 that once again underlines the basic features of process evolution discussed above. For more detailed analysis of the problem of hydrogen-oxygen ignition at the background of temperature gradient one can see Appendix in our previous papers [30, 64].

For the mixture compressed along the Hugoniot curve (behind the compression or shock waves) the burning velocity-pressure dependence can be of certain interest. However it should be noted that the pressure dependence is much weaker than the temperature dependence. On the other hand, the joint increase of pressure and temperature causes competition between the reaction mechanisms (one for low temperatures and high pressures and another one for high temperatures and low pressures). In turn this competition may change drastically the flame structure. At elevated pressures the flame thickness decreases and the coupling of the reaction and shock fronts becomes possible. For the case considered in the present paper a coupling of the reaction front and shock occurs at about 8-10atm and ambient temperature of about 600K. The flame front propagating in such conditions is thinner than one at normal conditions and resolved with about 20 computational cells. On the other hand the scheme diffusivity of the numerical method smoothens the shock front over 5 computational cells. Therefore no artificial coupling is possible for the chosen meshes determining fine resolution.

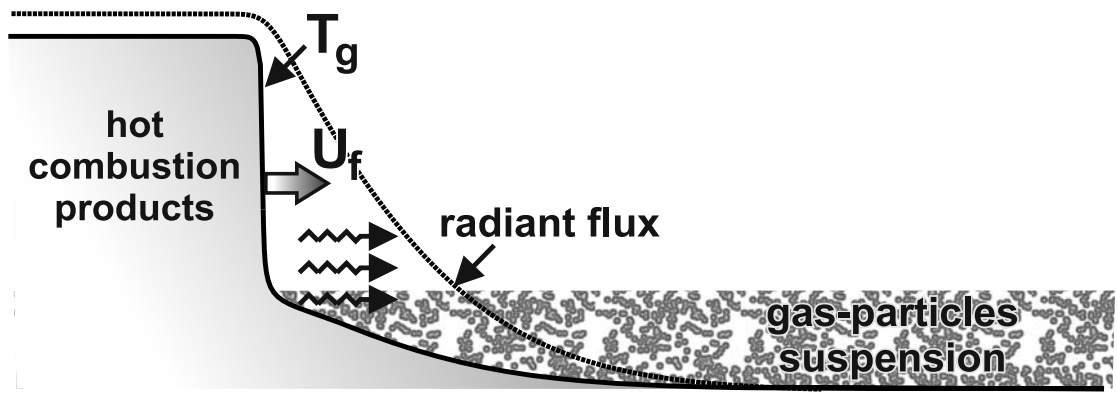

**Figure 1.** Scheme of the flame propagation through the gas-particles cloud (a gaseous mixture with uniformly suspended particles).

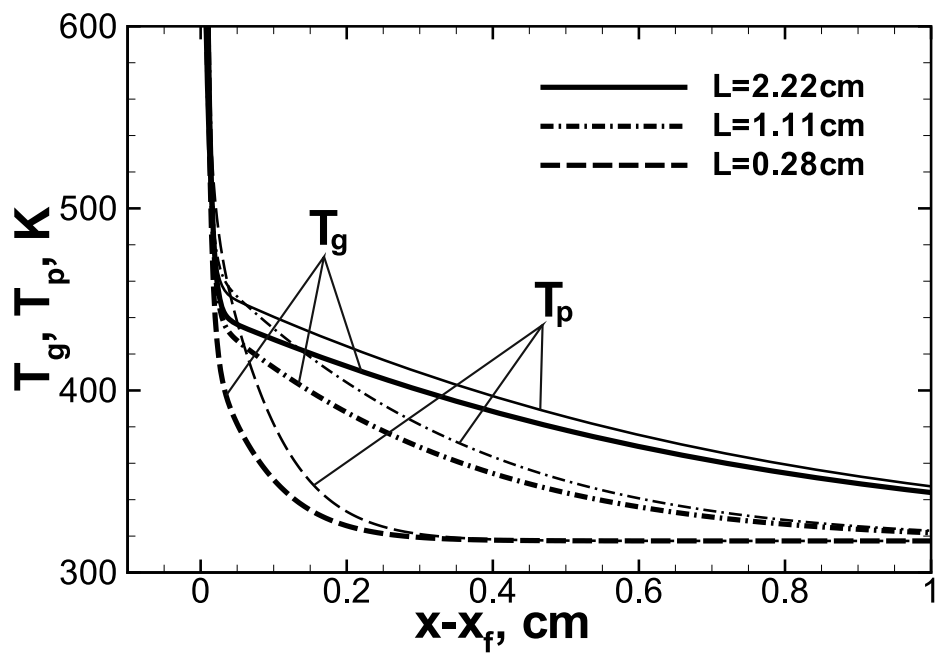

**Figure 2.** Temperature distribution ahead of the flame front calculated for uniformly distributed suspended particles and for different radiation absorption lengths, L=0.28, 1.11, 2.22 cm. Thick curves represent the gaseous temperature $T_g$, thin curves – particles' temperature $T_p$.

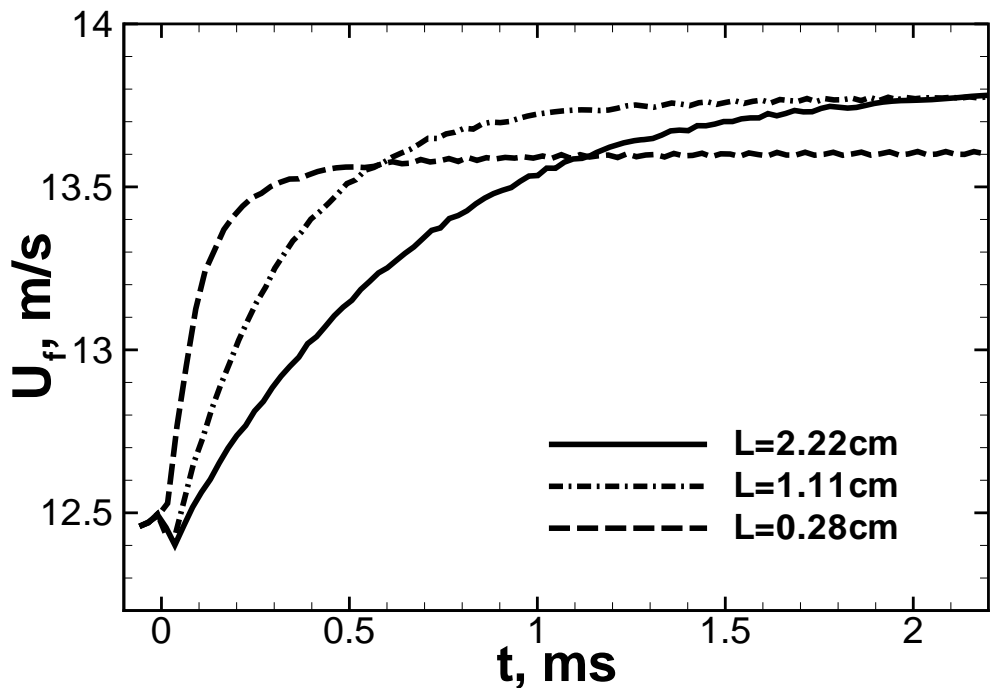

**Figure 3.** Burning velocity increase as the flame propagates through the gas-particles cloud with different thermal radiation absorption lengths (L).

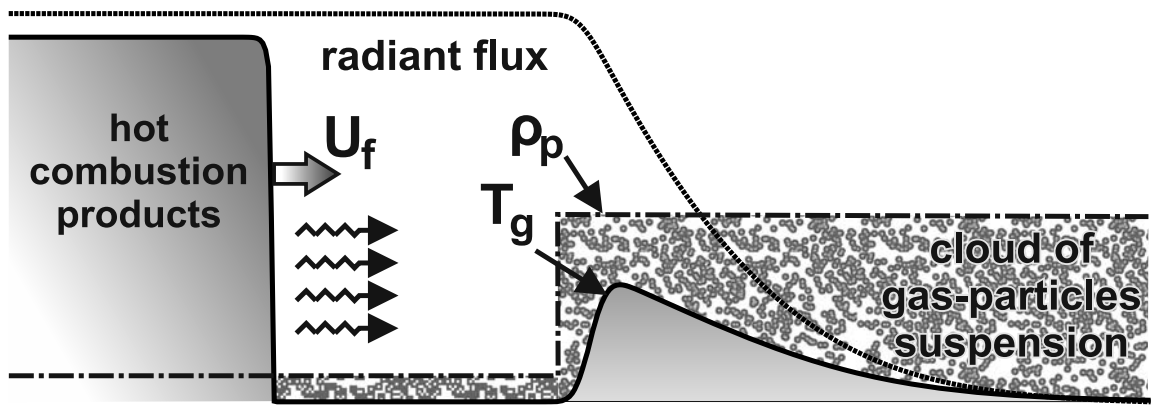

**Figure 4.** Scheme of the radiant preheating of the gaseous mixture inside the gas-particles cloud far ahead the flame front.

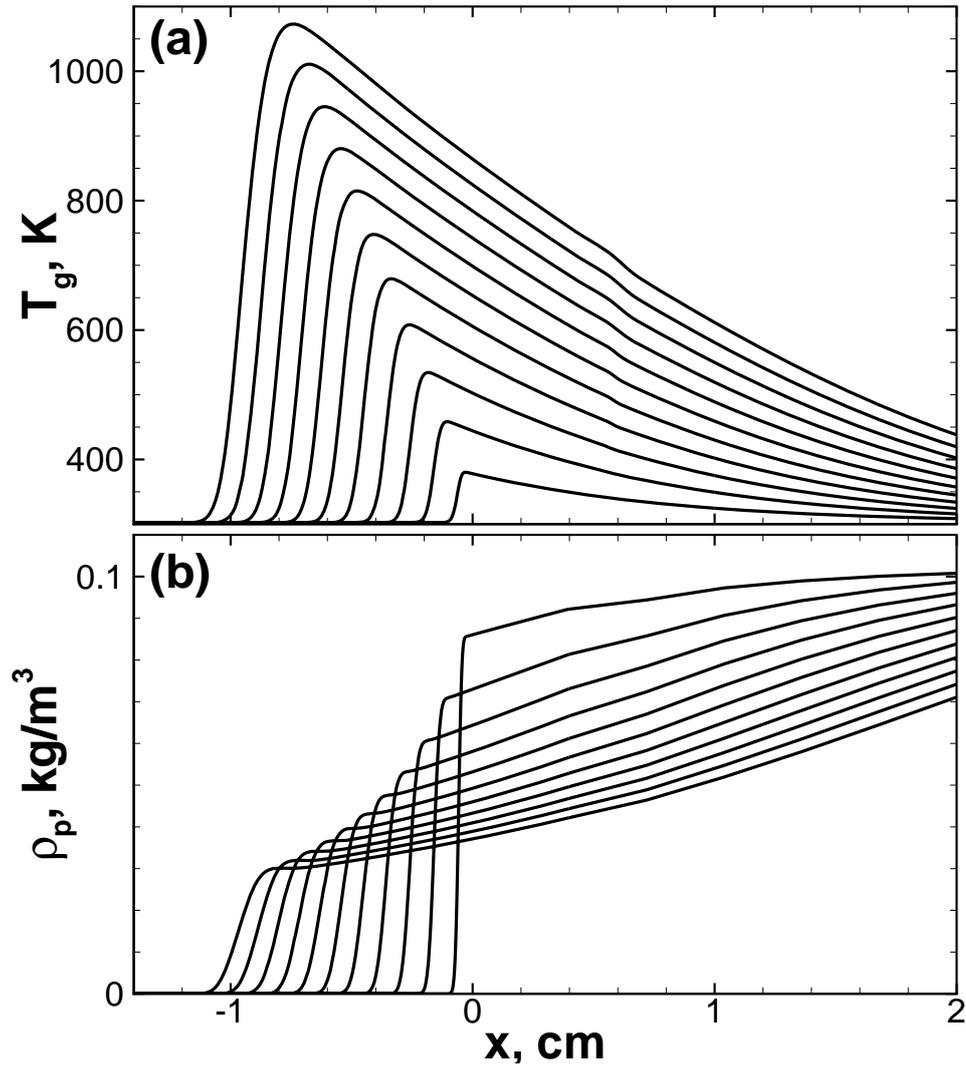

**Figure 5.** Time evolution of the gaseous temperature (a) and the mass density of the suspended solid particles (b) profiles during radiative preheating of the gas-particles cloud ahead of the advancing flame. Profiles are shown with the time intervals of 80μs. For initial stepwise particles density profile, $N_p = 2.5 \cdot 10^7 \, cm^{-3}$, $r_p = 1 \mu m$.

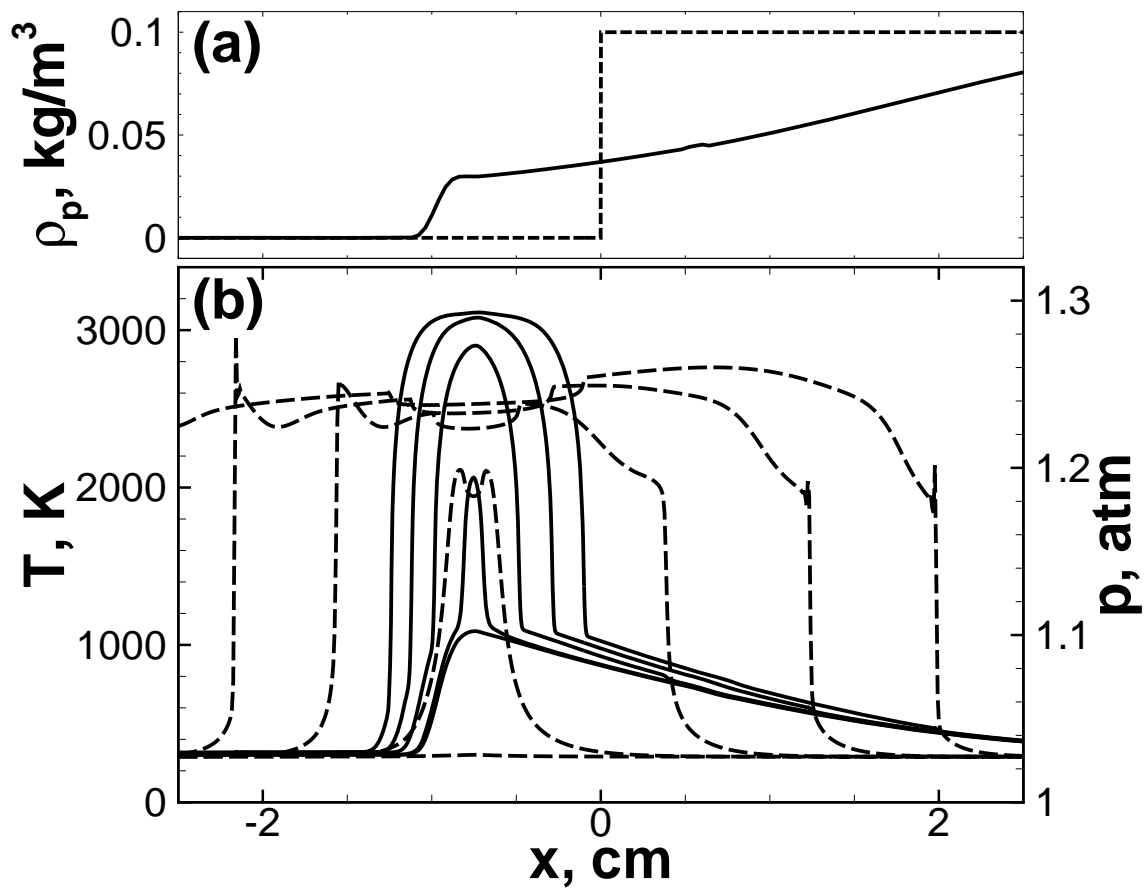

**Figure 6.** The upper frame (a) shows the distribution of particles mass density: dashed line – the initial stepwise density profile, solid line – density profile at time instant $t_0$ prior to the ignition. (b): Time evolution of the gaseous temperature (solid lines) and pressure profiles (dashed lines) during the formation of deflagration in the vicinity of the margin of the gas-particles cloud ~~far~~ ahead of the propagating flame. The profiles are shown at sequential time instants starting from $t_0=890\mu s$ with intervals $\Delta t=10\mu s$.

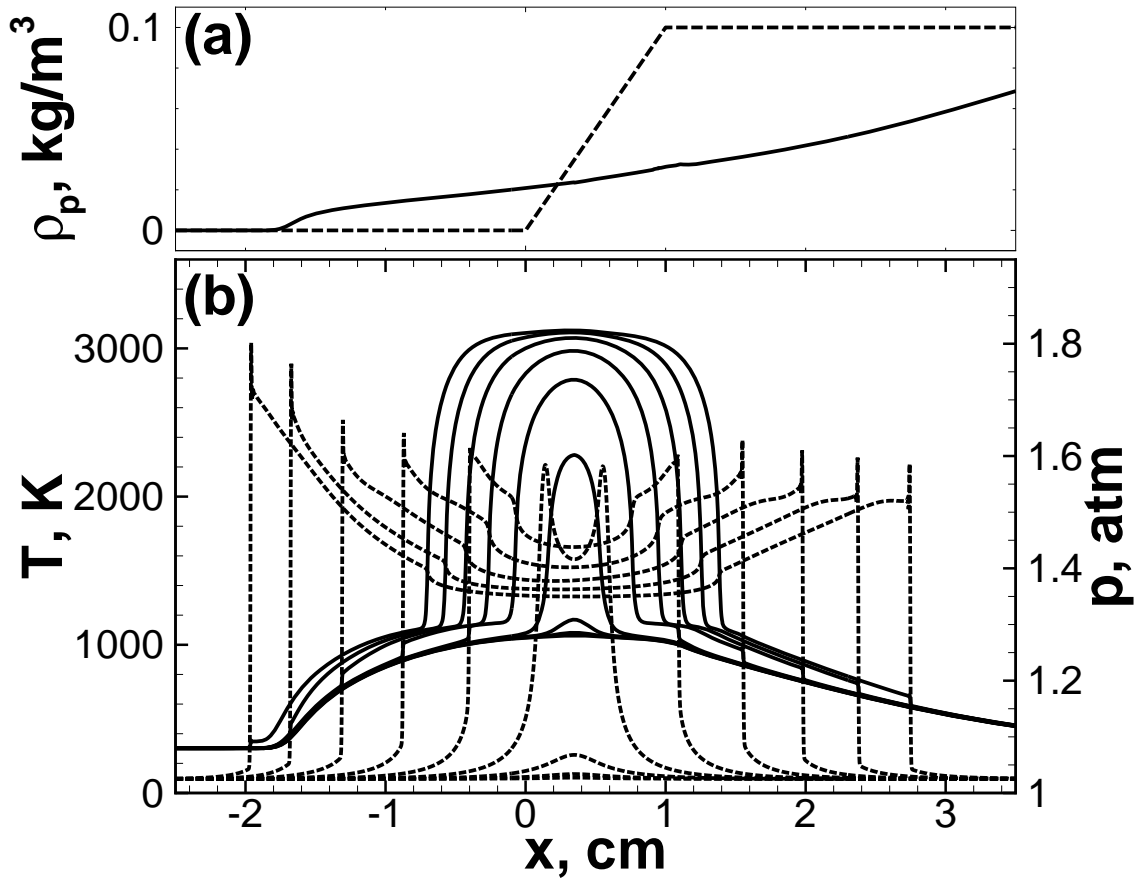

**Figure 7.** (a): The distribution of particles mass density: dashed line – the initial linear density profile of width 1.0cm, solid line – density profile at time instant $t_0$ prior to the ignition. (b): Time evolution of the gaseous temperature (solid lines) and pressure (dashed lines) profiles during the formation of fast deflagration wave behind the outrunning shock in the vicinity of the margin of the gas-particles cloud ahead of the advancing flame. The profiles are presented at sequential time instants starting from $t_0=1650\mu s$ with intervals $\Delta t=4\mu s$.

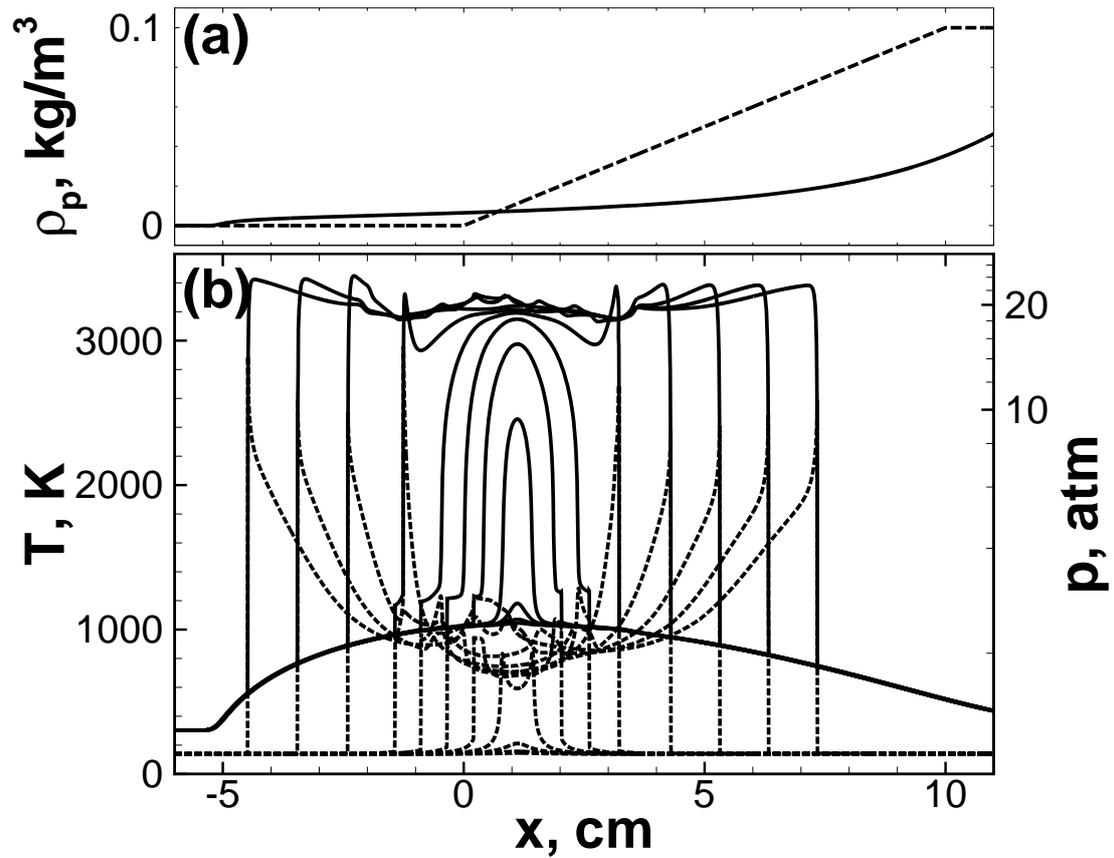

**Figure 8.** (a): The distribution of particles mass density: dashed line – the initial linear density profile of width 10cm, solid line – density profile at time instant $t_0$ prior to the ignition. (b): Time evolution of the gaseous temperature (solid lines) and pressure (dashed lines) profiles during the detonation formation in the vicinity of the left boundary of the gas-particles cloud ahead of the advancing flame. The profiles are presented at sequential time instants starting from $t_0=5015\mu s$ with intervals $\Delta t=4\mu s$.

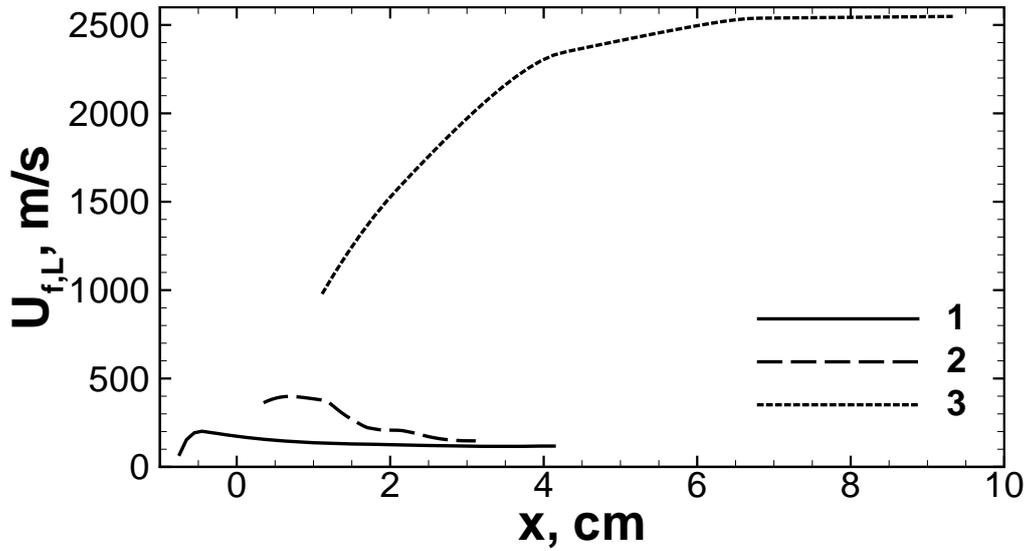

**Figure 9.** Evolution of the spontaneous reaction wave velocity along the temperature gradients during the formation of deflagration, fast deflagration and detonation (curves 1,2,3) presented in figures 6-8.

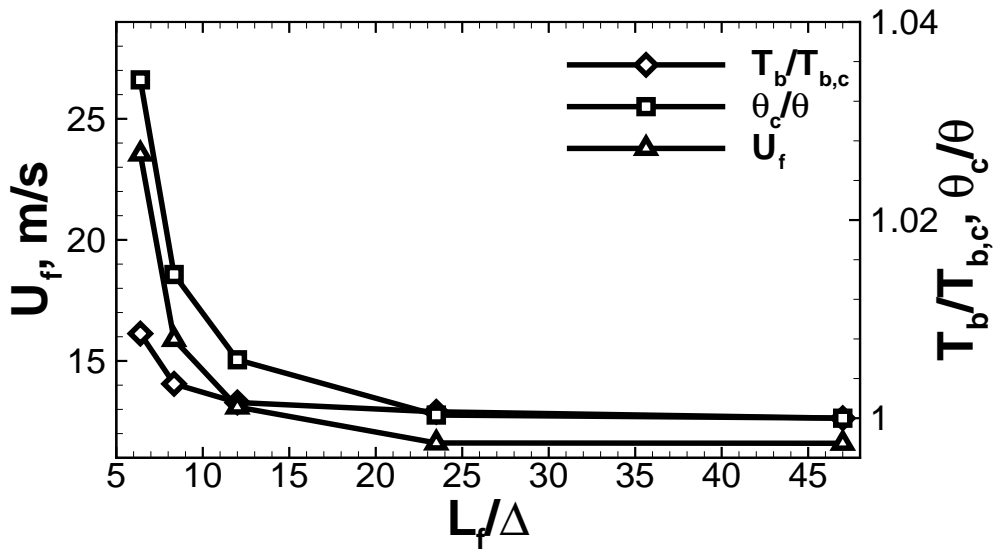

**Figure A1.** Resolution test for normal velocity of stoichiometric hydrogen-oxygen flame at normal ambient conditions ($T_0 = 300K$, $p_0 = 1atm$). $U_f$ – burning velocity, $T_b$ – adiabatic temperature of the combustion products, $\theta = \rho_u / \rho_b$ – expansion ratio, index 'c' responds to the converged value.

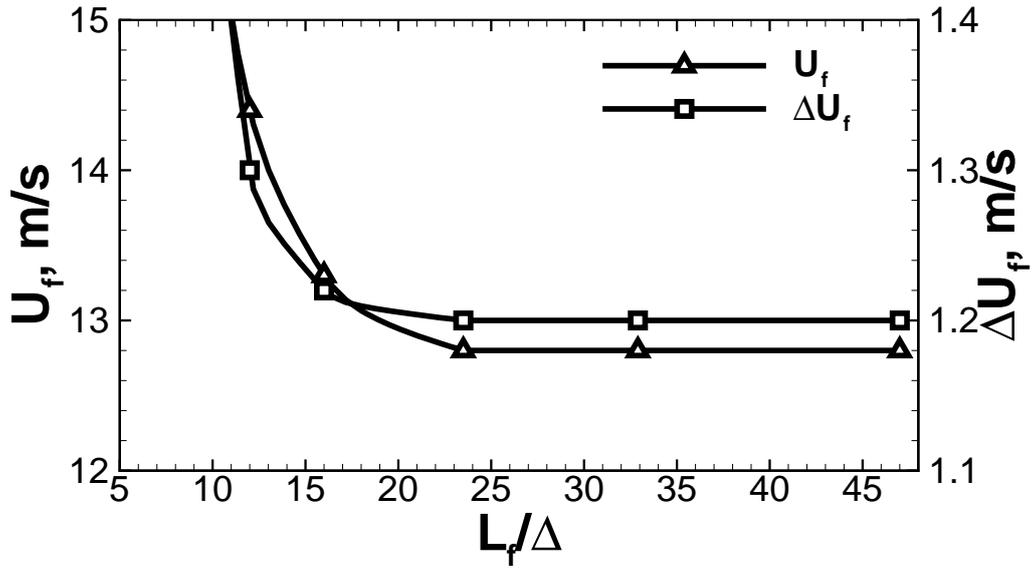

**Figure A2.** Resolution test for problem of the flame propagating in the uniformly seeded mixture, L=2.22cm. $U_f$ – burning velocity, $\Delta U_f$ – velocity increase due to the radiant preheating of the mixture ahead of the flame front.

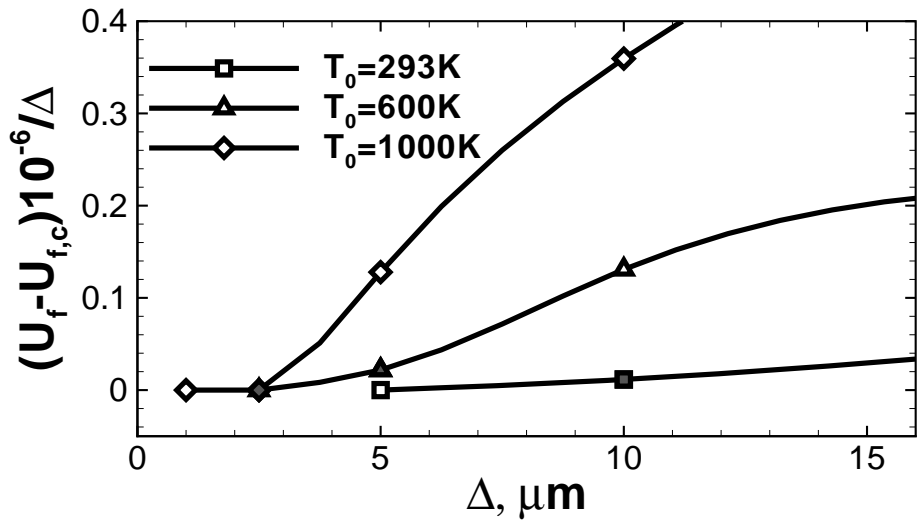

**Figure A3.** Resolution tests for different ambient temperatures. Grey signs show acceptable range of convergence. $U_f$ – is a normal flame velocity reproduced with cell size $\Delta$, $U_{f,c}$ – is a converged solution for $U_f$. The value $(U_f - U_{f,c})/\Delta$ shows the convergence rate at given resolution.

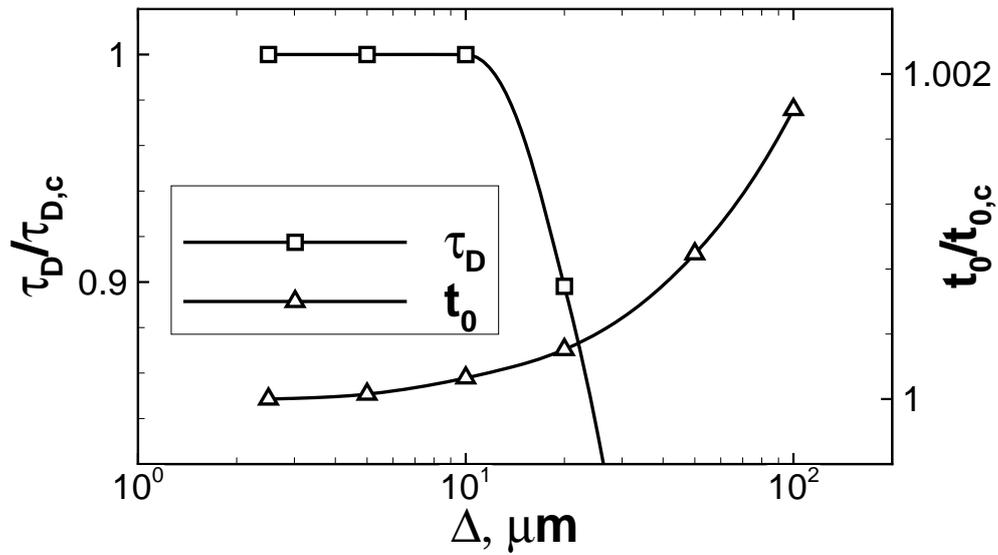

**Figure A4.** Resolution test for the detonation ignition in the non-uniformly seeded mixture, the problem setup corresponds to the solution presented in Figure 8. $t_0$ – preheating time, $\tau_D$ – duration of the detonation onset. Index 'c' responds to the converged value.

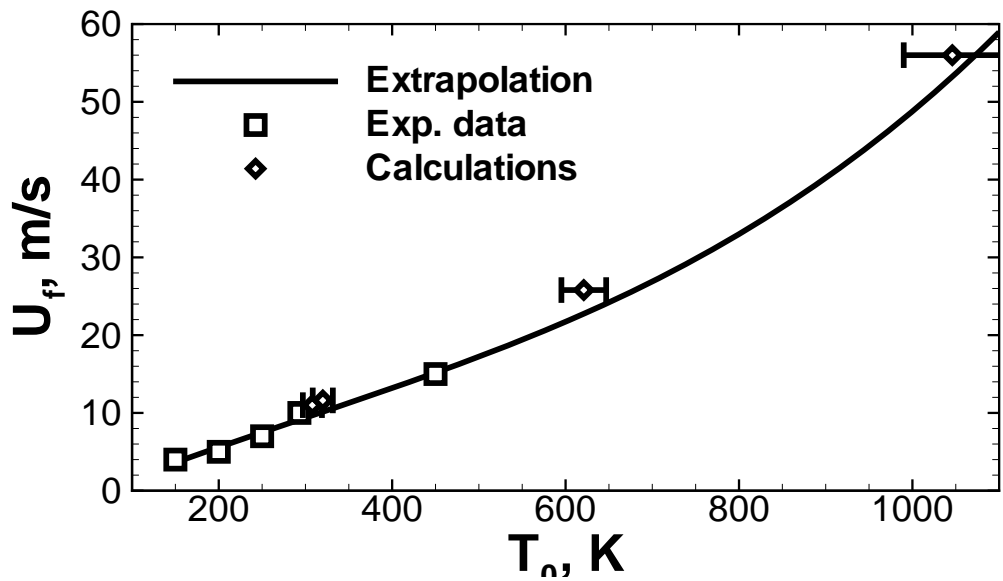

**Figure A5.** Stoichiometric hydrogen-oxygen burning velocities versus the ambient temperatures. Solid line shows the extrapolation of the experimental data obtained in [63] and signed with squares. Diamonds represent the calculated data with error bars determined by the temperature rise behind the compression wave emerged from the ignition kernel.